\documentclass[conference]{IEEEtran}

\usepackage[ruled]{algorithm2e}
\usepackage{stmaryrd}
\usepackage{graphicx}
\usepackage{color}
\usepackage{comment}
\usepackage{url}
\usepackage{wrapfig}
\usepackage{subfigure}
\usepackage{tabularx}

\newcommand{\superscript}[1]{\ensuremath{^{\textrm{#1}}}}

\def\wu{\superscript{*}}
\def\wg{\superscript{\dag}}
\def\wa{\superscript{\S}}

\begin{document}

%\title{A Model-Driven Communication for Distributed Medical Best Practice Guidance Systems}
\title{A Pathophysiological Model-Driven Communication for Dynamic Distributed Medical Best Practice Guidance Systems}

\author{\IEEEauthorblockN{Mohammad Hosseini\wu, Yu Jiang\wu, Poliang Wu\wu, Richard B. Berlin Jr.\wu\wg, Shangping Ren\wa, Lui Sha\wu}
\IEEEauthorblockA{
%  \sharedaffiliation
  \begin{tabular}{ccc}
    \wu Department of Computer Science &  \wg Department of Surgery & \wa Department of Computer Science\\
    University of Illinois at Urbana-Champaign    &  Carle Foundation Hospital 	&	 Illinois Institute of Technology\\
  \end{tabular}
  ~\\
Email: \{shossen2, jy1989, wu87, rberlin, lrs\}@illinois.edu, ren@iit.edu}}

\maketitle

\begin{abstract}
There is a great divide between rural and urban areas, particularly in medical emergency care. %The highest death rates are found in rural counties. For emergency patient care, time to definitive treatment is critical. However, deciding the most effective care for an acute patient requires knowledge and experience. 
Although medical best practice guidelines exist and are in hospital handbooks, they are often lengthy and difficult to apply clinically. The challenges are exaggerated for doctors in rural areas and emergency medical technicians (EMT) during patient transport.

In this paper, we propose the concept of distributed executable medical best practice guidance systems to assist adherence to best practice from the time that a patient first presents at a rural hospital, through diagnosis and ambulance transfer to arrival and treatment at a regional tertiary hospital center. %A patient in a rural hospital with signs and symptoms suggestive of stroke is used for illustrative purposes. 
We codify complex medical knowledge in the form of simplified distributed executable disease automata, from the thin automata at rural hospitals to the rich automata in the regional center hospitals. However, a main challenge is how to efficiently and safely synchronize distributed best practice models as the communication among medical facilities, devices, and professionals generates a large number of messages. This complex problem of patient diagnosis and transport from rural to center facility is also fraught with many uncertainties and changes resulting in a high degree of dynamism. A critically ill patient's medical conditions can change abruptly in addition to changes in the wireless bandwidth during the ambulance transfer. Such dynamics have yet to be addressed in existing literature on telemedicine.
%Due to the uncertain nature of patient care and physiologic response as well as changes in the patient condition, available diagnoses and treatment options that occur, the inherent distribution of communication needs incurs a high degree of dynamism. This often causes the existing distributed communication mechanisms to fall short and inadequately support synchronization requirements. % because of their \textit{static} communication semantics. 
To address this situation, we propose a pathophysiological model-driven message exchange communication architecture that ensures the real-time and dynamic requirements of synchronization among distributed emergency best practice models are met in a \textit{reliable} and \textit{safe} manner. Taking the signs, symptoms, and progress of stroke patients transported across a geographically distributed healthcare network as the motivating use case, we implement our communication system and apply it to our developed best practice automata using laboratory simulations. Our proof-of-concept experiments shows there is potential for the use of our system in a wide variety of domains.
\end{abstract}
\IEEEpeerreviewmaketitle
%\category {C.3} {Computer Systems Organization}{Special-Purpose and Application-Based Systems}
%\category {J.3} {Life and Medical Sciences}{}
%\keywords{Medical Best Practice Guidance Systems, stroke, Model-Drivel Emergency Communication}

%\newpage
\section*{Acknowledgment}
This paper reviews a technology package that is the result of a team effort. Lui Sha led the development of best practice medical guidance systems and all system integration. Richard Berlin conceived and described the problem and outlined the system approach for the study of acute stroke patient emergency care. Mohammad Hosseini, Shangping Ren, Poliang Wu, and Jiang Yu led the development of simplified stroke models, workflow management, and distribution. Mohammad Hosseini led the literature review, study, and development of communication architecture, protocol, and simulations that were conducted.

The material presented in this paper is based upon work supported in part by NSF CNS 1329886, by NSF CNS 1545002, and in part by ONR N00014-14-1-0717. Any opinions, findings, and conclusions or recommendations expressed in this publication are those of the authors and do not necessarily reflect the views of the grant providers.

\section{Introduction}
There is still a great disparity in medical care system support across large geographic regions, most profoundly for emergency care, where limited facilities and remote location play a central role. Based on the Wessels Living History Farm report~\cite{urban-rural-diff}, the doctor to patient ratio in the United States is 30 to 10,000 in large metropolitan areas, only 5 to 10,000 in most rural areas; and the highest death rates are often found in the most rural counties. Currently, more than 60 million Americans live in rural areas and face challenges in accessing high-quality medical care~\cite{rural-care-death-rate}. For emergency patient care, time to definitive treatment is critical. However, deciding the most effective care for an acute patient requires knowledge and experience as well as infrastructure support. Although medical best practice guidelines are accepted and widely available in hospital handbooks, such guidelines are often lengthy and difficult to apply clinically. The challenges are exaggerated for doctors in rural areas and emergency medical technicians (EMT) during patient transport.

%stroke is the third leading cause of death and the first leading cause of disability in the United States~\cite{stroke-stats}. In addition, stroke patients are often elderly (in fact, 65\% to 72\% stroke patients are over age 65~\cite{elderly-stroke-rate}) who may have other illness, such as heart diseases and diabetes. Furthermore, some effective stroke treatment medications have constraints. These factors not only call for new research to provide more effective acute stroke patient care, but also make it more challenging to provide computer and communication technology support for stroke patient care.

In this work, the team developed an advanced cyber-physical-human system technology to transform emergency care for acute patients in a hospital network covering a large rural area. %Clinically, our system is designed for many emergency cares including trauma, heart attack, strokes, and sepsis, as well as many situations other than care, such as those seen in disaster response scenarios \cite{dcoss14}. 
The technology enables the adherence to best practice guidelines from rural hospitals, during ambulance transfer, through arrival at the regional center hospital. Although applicable for many life-critical systems, we focus on many-faceted stroke symptomatology and presentation of illness as a motivating use case, and illustrate a scenario given a patient in a rural hospital with signs and symptoms of stroke. We codify complex medical knowledge in the form of simplified executable automata, and use them to propose \textit{dynamic} distributed emergency best practice models that can be shared between the rural hospitals on the ambulance, and at the regional center hospital. The best practice models are executed in real-time at both rural and center hospitals, with doctors in the center hospital supervising the rural hospital doctor as both follow best practice based on patient pathophysiological information that is simultaneously monitored at both locations and in the ambulance. However, despite the promising nature of adherence to best practice models, a challenge remains in the rural facility transport example in that the distribution requires that distributed executable models become synchronized to keep current states of distributed models consistent with each other.

This complex problem of patient diagnosis and transport from rural to center facility is fraught with many uncertainties and change in patient condition resulting in a high degree of dynamism. When a patient is being transferred from rural hospital to ambulance and then to a regional hospital center, the available diagnoses and treatment options vary greatly. And a critically ill patient's medical conditions can change abruptly in addition to changes in the wireless bandwidth during the ambulance transfer. The distribution of communication needs can adapt over time, and new automata can leave or join the best practice system in response to variations in physiological and physical conditions, such as capabilities, patient, or disease models. Such dynamics have yet to be addressed in existing literature on telemedicine as existing distributed communication mechanisms such as those used by Remote Procedure Call (RPC) or Remote Method Invocation (RMI) fall short to support synchronization requirements due to their \textit{static} communication semantics. Lack of support for dynamic invocations introduces fundamental implementation issues, such as handling failures at client and/or server using request-reply protocols, parameter passing, etc. \cite{dsp,dsc}. That said, existing tele-medicine technologies such as \cite{telemedicine1, telemedicine2, telemedicine3} provide healthcare communication via remote audio/visual monitoring of patients, with \textit{unstructured} and \textit{static} communication semantics. Our work on the contrary, is centered around the novel concept of distributed medical best-practice systems and the messaging aspect of tele-medicine introduced with structured communication and a high degree of dynamism, which is distinguished form the boundaries of existing tele-medicine systems.

To address the challenges that exist, we then propose a pathophysiological model-driven message exchange architecture for dynamic distributed best practice systems with the aim of synchronizing the distributed pathophysiological models of patients in rural and center hospitals given their timely changes, while discussing how it can meet the \textit{reliability} and \textit{safety} requirements of dynamic distributed emergency best practice systems. Our proposed communication architecture can be extended to implement the distributed best practices in emergency first response systems not directly related to medicine, such as those seen in disaster response scenarios \cite{dcoss14}. In summary, the main contributions of this paper include the following:
\begin{itemize}
\item The description of distributed medical best practice guidance systems and the dynamism introduced by this distribution, as well as the concept of synchronization among the distributed executable automata. This new generation of synchronized best practices enhances the overall treatment time for emergency care.  
\item Codification of medical best practice knowledge into executable automata. We take signs and symptoms suggestive of stroke as the motivating use case to conceptualize distributed medical best practice guidance systems.
\item Design of a dynamic message-exchange communication architecture built for synchronization of distributed safety-critical executable best-practice automata.
\end{itemize}

%To the best of our knowledge, this is the first work to propose the notion of \textit{``distributed medical best practice guidance system''} and the dynamism introduced by the distribution, \textit{``synchronized distributed executable automata''}, as well as \textit{``dynamic message-exchange architecture for emergency care''} built for synchronization of distributed safety-critical executable best-practice automata.

The remainder of the paper is organized as follows: In Section \ref{sec:background}, we provide background information illustrating a stroke emergency care scenario, and cover a wide area of related work about best practice systems, executable automata, and communication systems, while discussing how our work is related to them. In section \ref{sec:distribution} we discuss the notion of distributed medical best practice systems and the dynamism introduced by the distribution. In section \ref{sec:architecture}, we explain our methodology for the design of a message-exchange communication architecture, including discussion of registration procedures for architectural components, failure detection, as well as the safety features of the protocol. Our proof-of-concept simulation is discussed in Section \ref{sec:implementation}, while in Section \ref{sec:conclusion} we conclude the paper and briefly discuss our plan for future work.

\section{Background and Literature Review}
\label{sec:background}

\subsection{Real Use-Case: Acute Stroke Patient Emergency Care}
We target stroke patient as a use case to investigate and envision how an ideal distributed best practice system and its communication support may improve the acute patient care in remote and rural areas. Stroke is the third leading cause of death and the first leading cause of disability in the United States~\cite{stroke-stats}. In addition, stroke patients are often elderly (in fact, 65\% to 72\% stroke patients are over age 65~\cite{elderly-stroke-rate}) who may have other illness, such as heart diseases and diabetes. Furthermore, some effective stroke treatment medications have strict implementation guidelines. These factors not only call for new research to provide more effective acute stroke patient care, but also make it more challenging to provide computer and communication technology support for stroke patient care.

\begin{figure*}[!t]
\centering 
%\hspace{-.5cm}
\includegraphics[width=1.7\columnwidth]{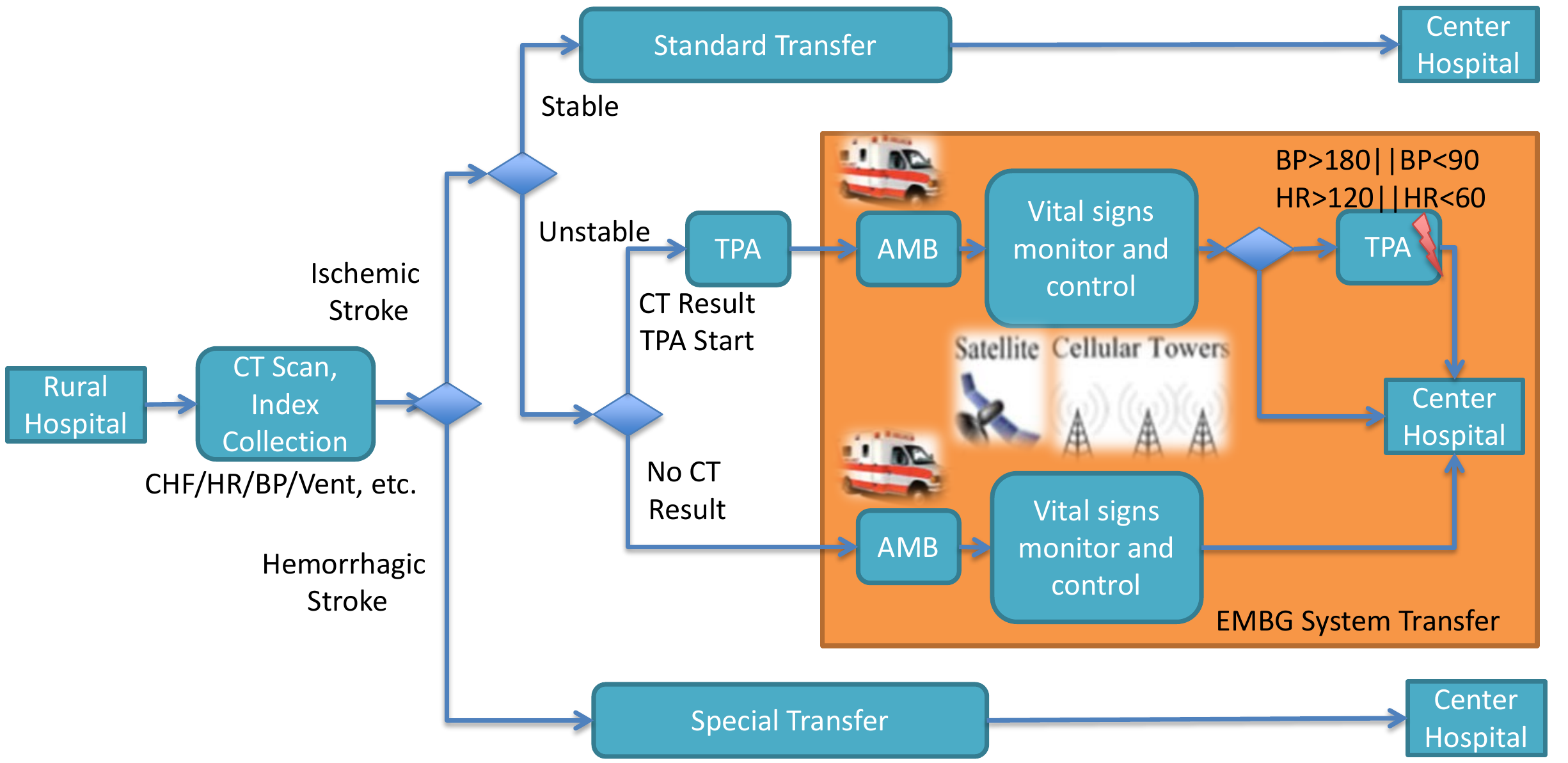} 
\caption{Envisioned workflow for stroke patients care from a rural to a center hospital}
\label{fig:example-tpa}
\end{figure*}

Figure \ref{fig:example-tpa} shows the envisioned workflow for stroke patients care from rural to center hospitals. Consider a 70 year old male patient arrives at a rural hospital. It is determined that the patient has the sudden onset of stroke. Computerized Tomography (CT) scan is completed and the images are sent to the stroke team at the center hospital for further immediate interpretation. His primary vital signs, measurements and indexes, such as blood pressure, heart rate, and blood oxygenation (SpO$_2$), etc., are promptly collected immediately. Patient's biometric information such as fingerprint can also be retrieved, sometimes using the camera mounted on a mobile device \cite{shervin1,shervin2,shervin3,shervin4,shervin5}, and sent if necessary. With the assistance from the regional center hospital via real-time monitoring (we collaborate with Carle Foundation Hospital as the center hospital \cite{carle}), the physicians and nurses in the rural hospital determine the patient's state and the types of the stroke.

If it is determined that the patient has a hemorrhagic stroke\footnote{Hemorrhagic stroke occurs when a blood vessel bursts inside the brain, which damages the nearby brain tissue.} (and not an ischemic stroke), he will be sent to the center hospital immediately because of the specials and specialized treatments that will be required. In the ambulance, supportive measures begin and the center hospital is notified and will be prepared for the patient. If the patient has ischemic stroke\footnote{Ischemic stroke occurs when a blood clot blocks an artery headed to the brain.} (much more common than a hemorrhagic stroke) and is in a stable state, standard actions such as tissue plasminogen activator (tPA) will be considered and administration begun; the patient will be sent to the center hospital with existing standard transfer approach through ambulance. Unfortunately, most patients with ischemic stroke are often in an unstable state, that is, vital signs are seriously out of range and must be treated actively. In this case of a patient with an ischemic stroke but unstable vital signs, the patient is placed on the ambulance accompanied with a stroke bag carrying needed equipment, blood products and treatments to manage care during transport and reduce the patient's risk of further deterioration.

tPA is a common treatment for blood clot which occludes an artery supplying blood to the brain and thus such clotting may be a common cause of ischemic stroke. A significant risk of the use of tPA to dissolve an occluding blood clot that has caused a stroke the complication of brain hemorrhage. The TEG\footnote{Thromboelastography (TEG) is a method of testing the efficiency of blood coagulation, which helps with timing tPA and clot dissolving therapy.} will be used to measure blood coagulation capability; the results will be sent to the center hospital to help guide the correct approach toward abnormalities in the stroke patient's blood clotting mechanism (many commonly taken medication have profound effects on blood clotting). In coordination with the expert consults at the central receiving hospital, the TEG results may also be used to address the infusion of tPA itself. In addition, patient's neurological symptoms such as speech difficulty, facial droop, weakness in hands and vital signs such as the blood pressure, heart rate, SpO$_2$, blood glucose level, and blood coagulation index will be monitored in real-time in the center hospital, at the rural hospital and the ambulance during transport. If any of the vital signs are out of range, the stroke team in the center hospital will work with rural physicians and the ambulance staff to manage treatment orders and the additional requirements that arise from a changing patient state.

For example, when the patient's blood pressure exceeds the safe threshold 180, the stroke team of the center hospital may suggest injecting the IV infusion of a medication such as nicardipine or nitroprusside to control the blood pressure. If the nicardipine or nitroprusside infusion does not control the elevated blood pressure or there are signs of further patient neurologic deterioration, the physician of the center hospital may change treatment accordingly. A blood glucose level that cannot be controlled within acceptable clinical range may be another measurement value which requires close coordination among the rural physician, the ambulance staff and the expert physicians at the central facility. Clearly, to minimize the patient's risks during transport, real-time supervision and monitoring of the stroke team in the ambulance is crucial, because the majority of stroke patients are elderly with various chronic diseases characterized by many complicating features. Even if such elderly patients are initially in a stable state, they may become unstable under the stress of stroke. Overall, the support of physicians' communication and real-time monitoring of patient and disease states is an important motivation in driving this research.

The design of our distributed best-practice system takes into consideration the different components that may impact the emergency care, and incurs real-time communication across rural hospitals, patient transport service, and center hospitals. The key components include (1) patient disease model, (2) facility models of rural hospitals, ambulances, and center hospitals, (3) patient conditions, and (4) any environment conditions such as road and traffic, weather, and communication coverage conditions. Overall, an effective emergency patient care involves an efficient, reliable, and safe communication and synchronization among all these models.

\subsection{Medical Best Practices for Emergency Care}
Medical best practice for emergency care have been created for patients in major hospitals~\cite{CMA, european2008guidelines, party2008national}. For instance, the University of Texas' MD Anderson Cancer Center has developed clinical management algorithms~\cite{CMA} that depict best practices for diagnosis, evaluation, and treatment of specific diseases including acute ischemic stroke targeting adult patients. While their contribution provides a high level algorithmic workflow using a multi-disciplinary approach, there are still several issues: a) their practice algorithms are specifically developed for MD Anderson Cancer Center and take into account circumstances particular to MD Anderson, including MD Anderson's specific patient population, their available services and structures, and its clinical information, b) the management algorithms are specially focused on those conditions that may arise during the course of cancer treatment, and c) their management algorithms lack enough details to handle patients who do not meet their necessary pre-requirements.

%Expert systems aim to capture domain-specific knowledge for inducing rules, making inferences and arriving at a specific conclusion~\cite{hudson2006medical}. Several inference and decision making methodologies have been developed, such as rule-based reasoning~\cite{tsumoto1996automated}, fuzzy logic~\cite{hudson1994fuzzy}, and probabilistic network~\cite{andreassen1991medical}. Medical expert systems are designed to provide medical diagnosis and treatment suggestions by codifying experts' knowledge, reasoning the logical rules and deducing new knowledge~\cite{tsumoto1996automated}. However, many clinical problems are complicated and involve dynamic decision-making, therefore straightforward attempts to chain together larger sets of rules unavoidably encounter challenges that are difficult to overcome~\cite{szolovits1988artificial}. Machine learning aims to excerpt common patterns from empirical medical data and make decisions based on the learned behaviors~\cite{kononenko2001machine}.

For stroke care for example, G\"{o}rlitz et al studied the feasibility of a stroke manager service concept using a combined service and software engineering approach, and developed workflow and IT architecture for improved post-stroke management~\cite{Gorlitz2012HPT}. Hofmann et al described concepts used for process optimization in stroke care and evaluated industrial methods to provide quality improvement in stroke management~\cite{Hofmann2012PM}. Panzarasa and Stefanelli likewise designed an evidence-based workflow management system as components of a knowledge management infrastructure by efficiently exploiting the available knowledge resources, aiming to increase the performance of higher quality of health-care delivery~\cite{Panzarasa2006NS}. While the proposed concepts in these studies consider requirements for enhanced stroke management, they were mainly positioned around health service networks, workflow, and knowledge management at individual health-care organizations. Many fundamental system and pathophysiological issues such as absence of stroke-specific expertise and high-end diagnosis and treatment equipment in rural areas, as well as problems associated with communication, distribution, and coordination among distributed best practices conditions still remain unaddressed. 

From the current practices, most notably, there is a need to develop novel \textit{executable}, \textit{distributed}, and \textit{dynamic} workflow automata that (1) adheres to best practices in acute patient care management, (2) ensures effectiveness and safety from both systematical and pathophysiological perspectives, and (3) enables efficient distribution of acute patient care across the rural area hospital, ambulance, and the center hospitals, given the many uncertainties and condition changes that exist. Unfortunately, the need has been neglected heretofore. It is worth pointing out that as a contribution to computer science and communications technology, we do not attempt to discover new medical knowledge. Rather, we focus on developing guidance system based on the accepted best practice medical system guidelines. The guidelines are high level instructions that are based on the pathophysiological models of patient's organ conditions. The complex interactions between organs in specific, provides the basis to codify disease dynamics in the form of interactive executable automata. %The diagnosis and treatment guidelines are different at each stage of an organ disease. Medical doctors are specialized organ by organ. The complex interactions between organs are often a source of medical errors. Fortunately, the guidelines are the results of a consensus of a team of leading specialists after years of research and clinical trials. This provides the basis to codify disease dynamics in the form of interactive executable automata.

\subsection{Executable Best Practice Automata}
\begin{figure}[!t]
\centering
\includegraphics[width=.7\columnwidth]{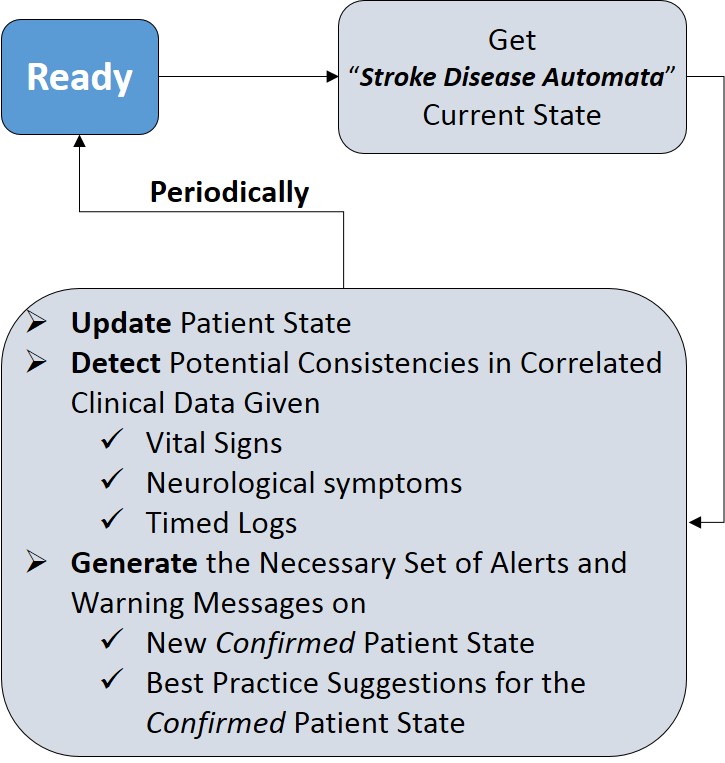} 
\caption{Stroke care manager abstract}
\label{fig:managerAbstract}
\end{figure}
From a medical perspective, physicians are taught organ system function as part of the representation of disease process. They look for patterns of pathophysiological changes (the change in physiological measurements as a result of disease) within an organ system to understand organ state~\cite{geary2010clinical, gao2007systematic}. This organ-centric view of pathophysiological expression also matches medical treatment, which is captured by best practice medical workflows. Therefore, the engine of our best practice systems is an executable best practice workflow model, and system automata such as disease or organ system automata. By codifying medical knowledge into executable formal best practice system automata, the codification can be checked by expert physicians via the execution of these models using scenario-driven simulation. %Once this is done, the desirable designs can be formally verified to ensure that they always hold.

In our previous work, we have proposed a Situation Awareness and Workflow Management (SAWM) system, and built best practice workflow and organ automata for cardiac arrest resuscitation \cite{maryam2014healthcare},%The architecture of the system is depicted in Figure \ref{fig:heartattack-fl}. 
with the states in each organ automaton, such as cardiac automaton, representing different organ states. The changes in the relevant physiological measurements and lab values which result in satisfaction of the condition for a new organ state causes state transitions. %When the attending physicians confirm the new organ states, the system then provides best-practice guidance based on the codified best practice workflow. 
In summary, SAWM system transforms passive text guidelines into a set of executable automata and helps physicians keep track of the states of automata. Based on the diagnosis from physicians, SAWM system provides step-by-step guidance in coherent with the workflows.

As a part of contributions in this paper, taking stroke as the motivating use-case, we codify medical knowledge into a simplified version of executable best practice workflow, and use that to develop a message-exchange architecture that can be used for communication and coordination among distributed best practice automata.

\subsection{Analogy: Space Communication Systems}
\label{sec:space}
%  \begin{wrapfigure}{R}{0.25\textwidth}
%\setlength{\columnsep}{2pt}%
%    \vspace{-.2cm}
%  \centering
%  \hspace{-.18cm}
%  \includegraphics[width=.5\columnwidth]{figure/managerAbstract.jpg}
 %   \caption{stroke care manager abstract} 
  %    \label{fig:managerAbstract}
%\end{wrapfigure}

Our medical best practice guidance system has an analogy with space communication systems, where a spacecraft communicates with the ground support system. In spacecraft control and fault recovery, a big brother copy is based on the ground station, which is fed with most recent information from spacecraft little brother to support the mission operation. The spacecraft does not need to run diagnostics, and therefore sends the information, called telemetry data, back to the ground system. Telemetry information is the data about the spacecraft needed to assess how well the spacecraft and the space mission operation are doing. Spacecraft attitude, power system measurements such as voltages of electronic systems on the spacecraft, the on/off status of all commandable equipment and heaters, as well as temperatures of components are examples of such data. For example, temperatures of key components are monitored on the ground station to make sure they do not overheat and malfunction. Should there be sudden rise of temperature, engineers on the ground station may decide to decrease utilization of the component, or other relevant systems. Loss of any major component on a spacecraft can affect all other components on a spacecraft and therefore, cause the mission to fail. The ground center copy is an exact model of the spacecraft, plus additional instrumentation that cannot be put into the spacecraft due to resource limitations, which in overall is considered a rich, extensive, and locally controlled model of spacecraft \cite{nasa1, boeing, voyager}.

Similar to space communication systems, for the medical best-practice guidance systems, a big brother model exists at regional center hospital which receives information from the little brother model located at rural hospital. The distributed best practice system is executed in real-time at both rural and center hospitals, with doctors in the center hospital supervising a rural hospital doctor to follow the best practices modeled by our guidance system. Figure \ref{fig:managerAbstract} shows an abstract overview of the care management process for stroke. Clinical information such as vital signs, neurological symptoms, and updates of disease states are monitored by doctors in the center hospital, potential consistencies are detected, and new patient state is then updated and confirmed. Next, appropriate best practice suggestions, corresponding warning messages, and necessary sets of alerts along with other invaluable information are generated, which are then being sent back to the rural hospital.

Current space communication systems such as NASA's Deep Space Network \cite{dsn1} involve protocols residing on all OSI layers, with some orthogonal aspects such as capabilities and security relying on services implemented at lower layers of the stack. However, unlike space communication systems, our design needs to be compliant with existing TCP/IP infrastructure. Furthermore, the problem is made more challenging by the fact that uncertainty of resources and patient heterogeneity as well as uncertainty resulting from human-in-loop nature of medical reasoning and evidence-based patient-centered care induces a high degree of dynamism to the emergency care communication system \cite{uncertainty1}, therefore requiring novel design considerations.

\section{Distributed Medical Best Practice Systems}
\label{sec:distribution}

\begin{figure*}[t]
  \centering
  \includegraphics[width=1.4\columnwidth]{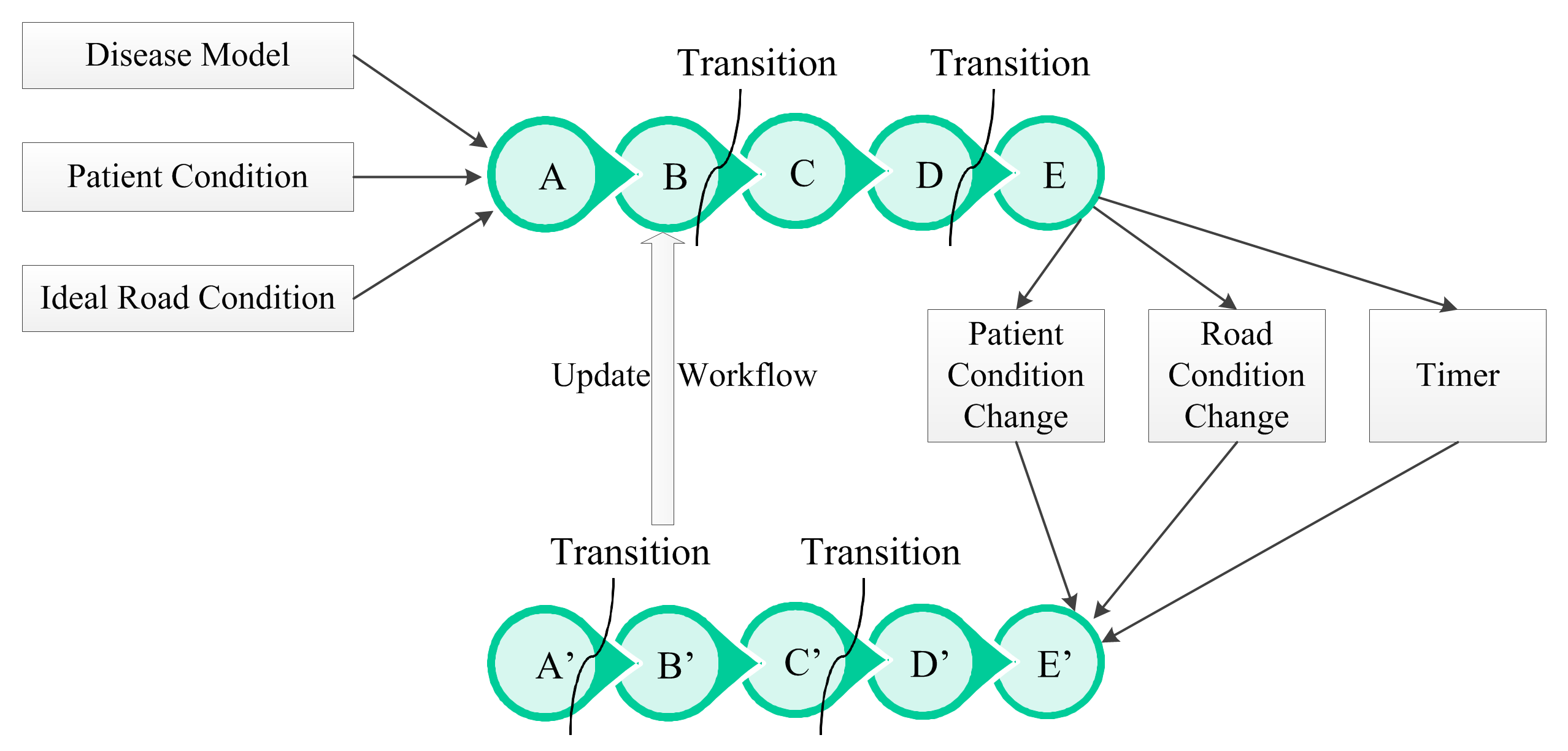} 
  \caption{Workflow adaption in dynamic distribution}
  \label{fig:wf-adaptation}
  \end{figure*}
  
The executable workflow automata as described in previous section focus on adherence to the best medical practice guidelines. They are based on various models, such as disease models, patient condition models, and models of facilities capabilities. To virtually extend a regional center hospital to its rural boundaries, the automata designed for different models have to be integrated together to form smooth and seamless care from rural to center hospitals. In addition, the distributed automata takes into consideration the physical environment between rural and center hospitals, such as communication coverage, weather, road, and traffic conditions as patient transport can be significantly impacted by the environment. What further exacerbates the distributed workflow challenge is the dynamic nature of distribution as system components such as  physical environment and patient condition can change rapidly. Both physical environment change and patient condition change can cause the workflow to be re-distributed or even a different automata to be introduced to the system and get activated. In stroke for instance, the scenario where a patient whose blood pressure increases above 180 after tPA treatment is begun will activate a new disease automata based on blood pressure control and change the automata of patient transport en route from one with best communication coverage for monitoring and consultation purposes to one that emphasizes fast transport of a patient whose conditions may require tPA to be discontinued pending blood pressure control or other complications. The transition at the center hospital procedure would be different as well if the patient situation is changed from initial ischemic stroke to hemorrhage stroke and should be able to continue implementing best practice accordingly and seamlessly.

\begin{figure}[!t]
\centering
\subfigure[Rural Hospitals]{
    \centering
    \includegraphics[width=\columnwidth]{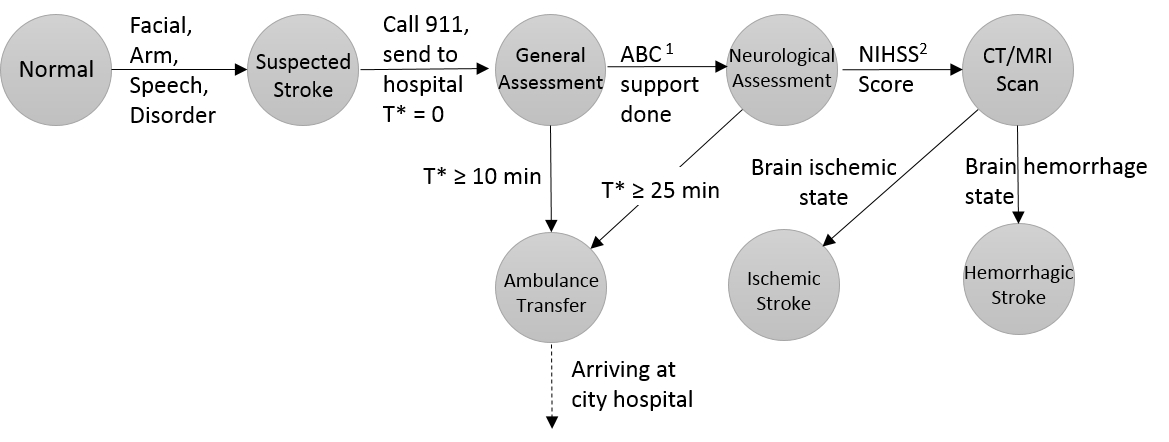} 
	\label{fig:strokeAutomataRural}}
\subfigure[Center Hospitals]{
	\centering 
	\includegraphics[width=\columnwidth]{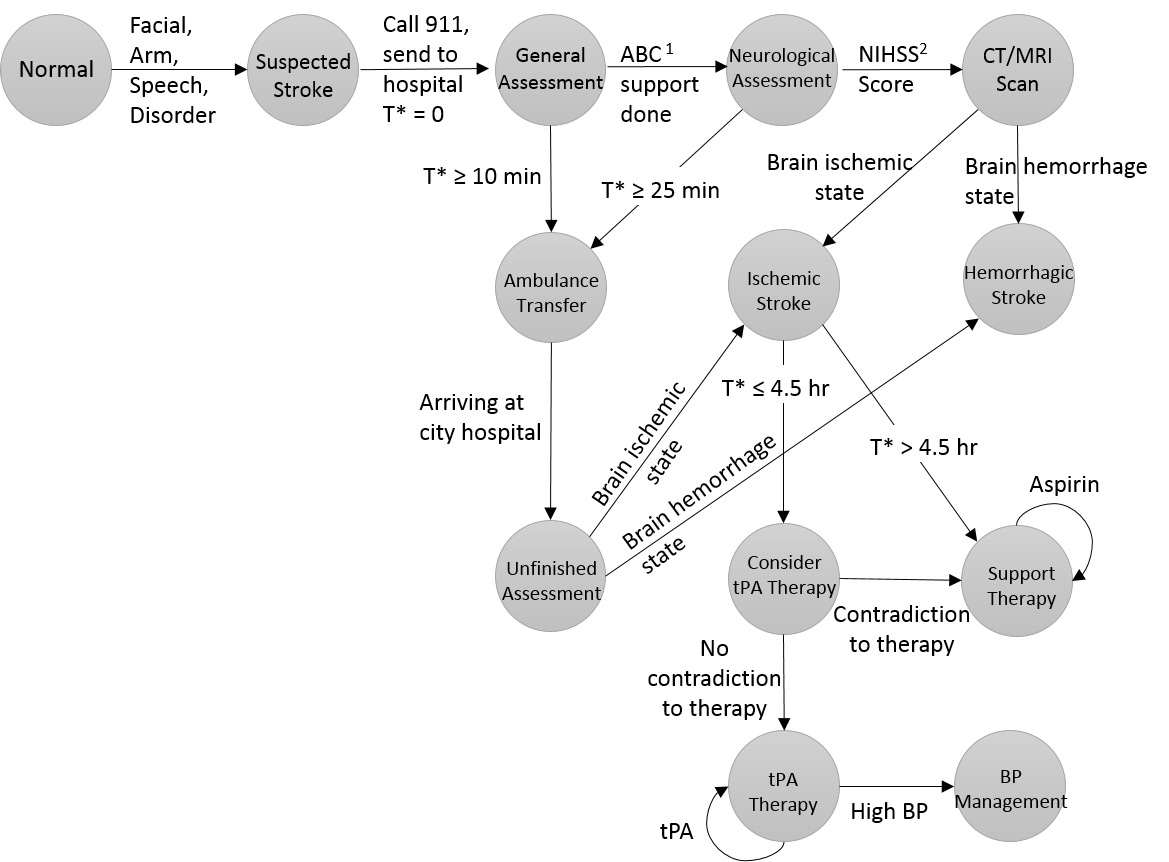} 
	\label{fig:strokeAutomataCenter}}
	%\vspace{-.5cm}
\caption{An instance of simplified distributed stroke automata}
\label{fig:strokeAutomata}
\end{figure}

The design of such distributed best practice system takes the responsibility and capability of rural and center hospitals and transport vehicles into consideration, and provides different levels of abstractions accordingly. In addition, as the initial diagnosis and performed treatments are automatically recorded, the best-practice system can continue the workflow automata after a patient is transferred to a center hospital, and help physicians and nurses at the center hospital seamlessly resume the patient treatment.

As what needs to be done and at what location are impacted by patient disease and patient current progress, capabilities of rural, ambulance, and center hospitals, and physical environment (e.g. communication coverage, weather, road, and traffic condition), the automata is first distributed based on static information, such as disease model and individual rural and center hospital capabilities, and then iterative and timely adaptation is made based on any changes in patient or physical conditions which overall forms a dynamic distribution. Figure ~\ref{fig:wf-adaptation} depicts the iterative procedure in forming a dynamic distributed acute patient care. %Moreover, in order to guarantee dynamic workflow adaptation is safely performed, safety requirements for workflow adaptation should be validated. In our previous work~\cite{WuTreatment2014, WuWorkflow2015}, we have developed treatment and workflow validation protocols. The developed protocols utilize pathophysiological models and validate the preconditions of treatments and workflows in collaboration with medical staff. It this work, we assume that the validation protocols will be integrated into our communication architecture to safely adapt workflows to disease stages and patient conditions.

Taking stroke as our use case, Figure \ref{fig:strokeAutomata} shows two greatly simplified stroke automata that are executed at a rural hospital (Figure \ref{fig:strokeAutomataRural}) and at a supervising regional center hospital (Figure \ref{fig:strokeAutomataCenter}). The figures represent an instance of a possible distribution since as mentioned before, variances in capabilities, expertise, and physical environment can cause dynamism in the sense of different cut-offs and different levels of abstraction for distributed executable workflow automata. The executable stroke automata in center hospitals as seen in Figure \ref{fig:strokeAutomataCenter} provide a rich, extended level of complications compared to the thin counterpart automata in rural hospitals as represented in Figure \ref{fig:strokeAutomataRural}. Many actions that can not be performed in rural hospitals due to lack of capabilities, such as supporting therapy using Aspirin, can only be performed in center hospitals as shown in Figure \ref{fig:strokeAutomataCenter}. However, it is always possible that thin models at rural hospitals have exclusive private states not common with the ones at center hospitals. Without loss of generality, we hereby assume that the simplified models at rural hospitals are a proper subset of the rich models at center hospitals. Overall, having the best practice models displayed in real-time at distributed locations, with doctors at center hospitals supervising a rural hospital by sending best-practice commands, will make the assistance much easier. %When a transferred patient arrives at the center hospital, our best-practice system continues executing the workflow given the distribution cut-off. Since the system tracks what has been done and the patient's response in real-time, it provides invaluable information to the physician at the center hospital.

As a major challenge however, development of a communication system among these dynamically distributed medical best-practice systems asks for a pathophysiological model-driven message-exchange architecture with a set of \textit{performance-wise} requirements that we uncovered during our case study. The message-exchange communication architecture shall: a) be efficient in the sense that it must meet the dynamic and real-time requirements of distributed emergency care, with best practice medical system components joining and leaving the system, b) scale to large numbers of messages and communicating automata, c) support priority in the communication protocol representing urgency of medical messages, d) be reliable in the sense that it must detect failure and monitor health of the architectural components due to the life-critical nature of medical best practice systems, and e) be safe in the sense that it avoids medical hazards in case communication fails. In the following, we will propose and discuss our pathophysiological model-driven message-exchange architecture.

\section{Message Exchange Architecture}
\label{sec:architecture}
 
  \begin{figure*}[!ht]
  \centering
  \includegraphics[width=\textwidth ]{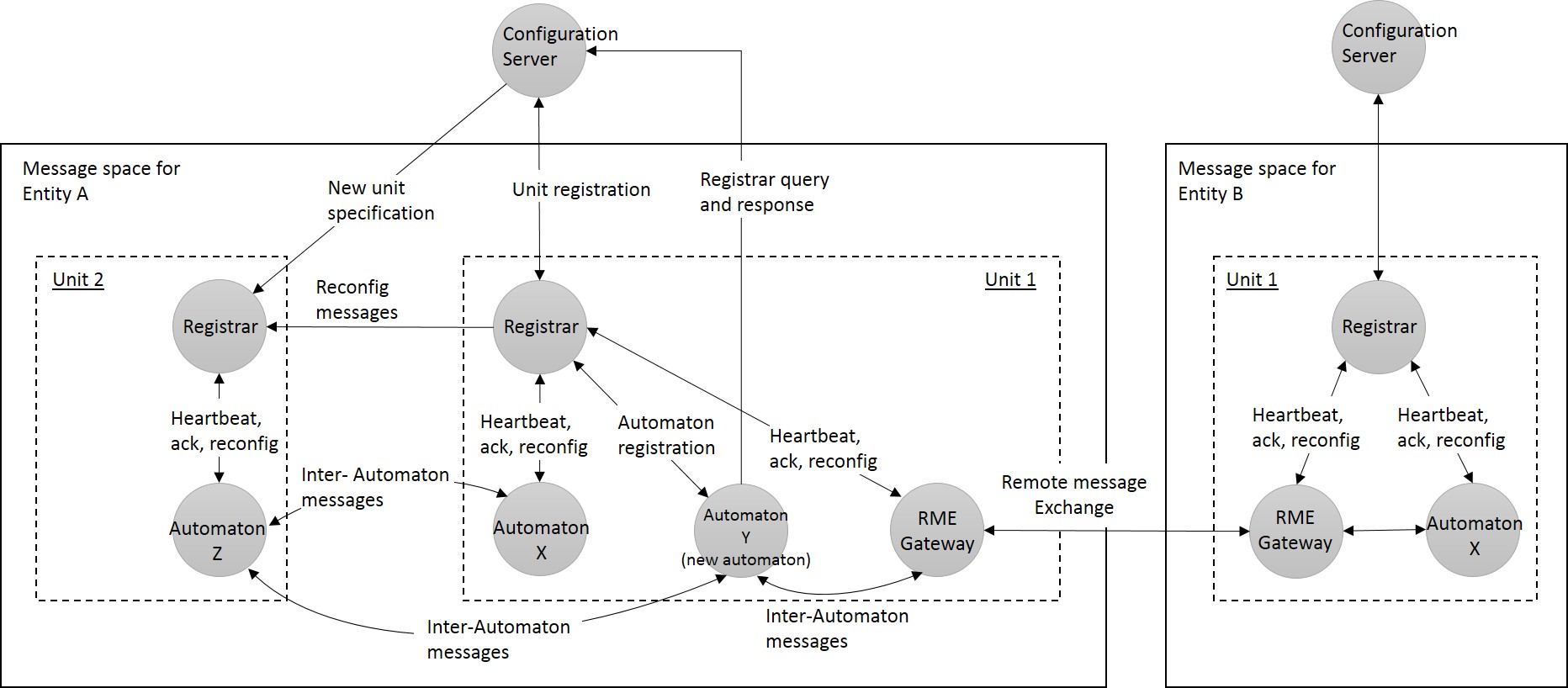} 
  \caption{Visual overview of the proposed message exchange architecture}
  \label{fig:architecture}
  \end{figure*}

Distributed best-practice models communicate by passing large number of various types of messages. A resulting key challenge therefore is how to efficiently synchronize the distributed executable best-practice automata from rural hospitals or the ambulance with the ones in the regional center hospitals. We borrow concepts employed in space communication systems such as \cite{dsn1}, \cite{ccsds1} and \cite{dsn2} to design a novel model-driven message-exchange communication architecture for dynamic distributed best practice automata, and apply significant addons and adjustments to meet our pathophysiological requirements and address the challenges discussed in Section \ref{sec:space}.

The main design goal of our message exchange communication architecture is to allow a dynamic model-driven message-oriented communication with reliability and safety requirements among various distributed best practice automata. These automata can simply represent any distinct workflow, such as disease automata, patient automata, or any other executable automata that may all reside in a single system or single location, or may inter-operate in a distributed system as seen in our distributed medical best practice system. Multiple instances of the same automata can operate concurrently in the same message space environment; they are distinguished by different hierarchical and layered UIDs with different authorities concerning their configuration and operation. Figure \ref{fig:architecture} shows an overview of the components of our proposed message exchange communication architecture. 

In our proposed architecture, the best-practice automata are managed by a registration server or registrar, that controls automata, monitors their status, transfers and receives configuration data. Every single automaton can only begin operation by announcing itself to a registrar. It learns the configuration information of its registrar such as its location or UID by querying a configuration server, which is responsible for monitoring and tracking the health of all registrars in any given location. 

We employ the notion of an entity, each consisting of multiple automata, registrars, and a configuration server that use message passing for the purposes of communication among themselves. Entity can represent executable models couples together at a specific location, such as rural or center hospitals, or an ambulance. While our communication architecture enables an entity to provide internal message-oriented communication, it can also communicate with an external entity remotely, through Remote Message Exchange (RME) gateways that we will describe in Section \ref{RME}. All message data are encrypted with the AES 128-bit symmetric cipher in electronic codebook (ECB) mode. To tackle complications made by single point of failure, a single entity can include multiple redundant configuration servers through a hierarchical ranking system. However, only a single, highest-ranked configuration server instance will operate at any given time.

An entity can be organizationally subdivided into units, or grouped automata, which are a group of role-related automata that overall make a consistent model. For example, an entity may consist of multiple automata such as executable disease automata, patient automata, or communication coverage automata. Each of these automata represents a single unit, which is categorized by the role it is designated to perform in the overall best practice system. Given this hierarchy, a unit is then consisted of a registrar and its associated automata. The subset of a units are automatons of that unit. For example, disease automata may include multiple organ automatons. In our previous work \cite{cardiac1}, we modeled cardiac arrest resuscitation disease as a combination of three organ automata, i.e. cardiac automaton, pulmonary automaton, and kidney automaton. Employment of such hierarchical sub-component-based architecture foster construction of delimited identifiers that helps an exchanged message to be uniquely transmitted among entities, and get directed to the targeted automaton. The design of our hierarchical message exchange communication architecture is independent of underlying transport and network protocols, and can therefore operate virtually on top of any communication methods, including existing TCP/IP infrastructure as well as the emerging Named-Data Networks (NDN). It can best facilitate delivery of messages between hierarchical sets of entities using a layered content descriptor naming design as promoted by NDN \cite{ccn1, ccn2, ccn3}. Our communication architecture also supports different message-oriented patterns such as send-receive, request-response, push-pull, publish-subscribe, is compliant with both synchronous and asynchronous communication, and can employ any combinations of these patterns to meet various application-specific requirements. %Designers alternatively may implement compliant systems focusing on a subset of these available capabilities, so to be able to meet various application-specific goals.

Due to the asynchronous nature of proposed message exchange communication architecture, a message does not incur a busy-wait suspension on an issuing automaton until a reply message is returned. That enables a high degree of concurrency in the overall performance. However, some key message exchanges naturally occur synchronously. For instance, as we will describe later a newly registering and initializing automaton must remain on busy waiting for responses form the configuration server and the registrar before running. In the future, we plan to apply the Physically-Asynchronous Logically-Synchronous (PALS) \cite{pals} pattern that we have developed to our proposed communication architecture, and investigate the trade-offs between the formal verification time and performance. %Therefore, our message exchange architecture is most suitable for contexts generally identified as networks with low and predictable latency, such as Internet or a stand-alone local area network.
 
In the followings, we describe an overview of configuration and registration process for major components, associated messages as well as general flow of messages, and introduce methods to detect failure and to enhance safety and reliability of the whole communication system.
\subsection{Initialization and Registration Process}
\subsubsection{Registrar Initialization and Registration}
  
  \begin{figure}[!t]
  \centering
  \includegraphics[width=.9\columnwidth]{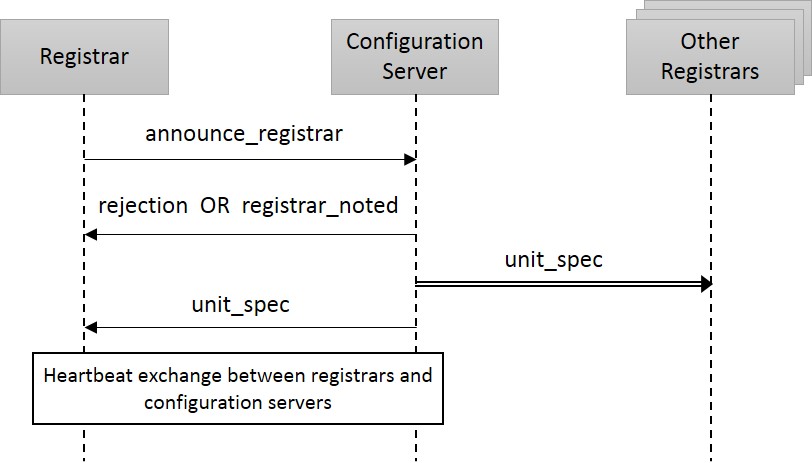} 
  \caption{Registrar initialization and registration process}
  \label{fig:registrar-initialization}
  \end{figure}
  
As mentioned previously, registrar is the main communication component in each unit that propagates information, monitors status, and acts as a registration server for every registered unit of multiple automata. Figure \ref{fig:registrar-initialization} shows registrar's initialization and registration process. At the beginning, each registrar is initialized by locating the configuration server and sends a ``announce-registrar'' configuration message to the configuration server. It is the responsibility of the configuration server to validate and verify the ``announce-registrar'' messages, so to make sure that the registrar belongs to a valid unit, and that the corresponding unit is not already initialized (given that no registrar is already active for that unit). The configuration server will then reply to the received message accordingly. Given that the validation and verification process is successful or not, two types of messages can be generated. If the validation process is failed, the configuration server generates a ``rejection'' message, and sends it back to the registrar. In case the validation process was successful, a ``registrar-noted'' message is generated which is sent back to the registrar. Upon a successful validation once the ``registrar-noted'' message is transmitted, configuration server will then generate a ``unit-spec'' message representing the updated status of the corresponding registered registrar, and sends it to all other registrars in the entity. In case the initialized registrar is the first and the only registered registrar, and there is no other active registrar, a single ``unit-spec'' message will be generated that will be returned to the corresponding registrar, with the message containing a unit ID set as the unique ID of registrar's unit number. Otherwise, one ``unit-spec'' message is generated and sent back to every other registrar in other units. From this point on, heartbeat messages are being exchanged between the registered unit and the configuration server periodically to detect possible failures and monitor unit's availability.

\subsubsection{Automaton Registration}

\begin{figure}[!t]
  \centering
  \includegraphics[width=\columnwidth]{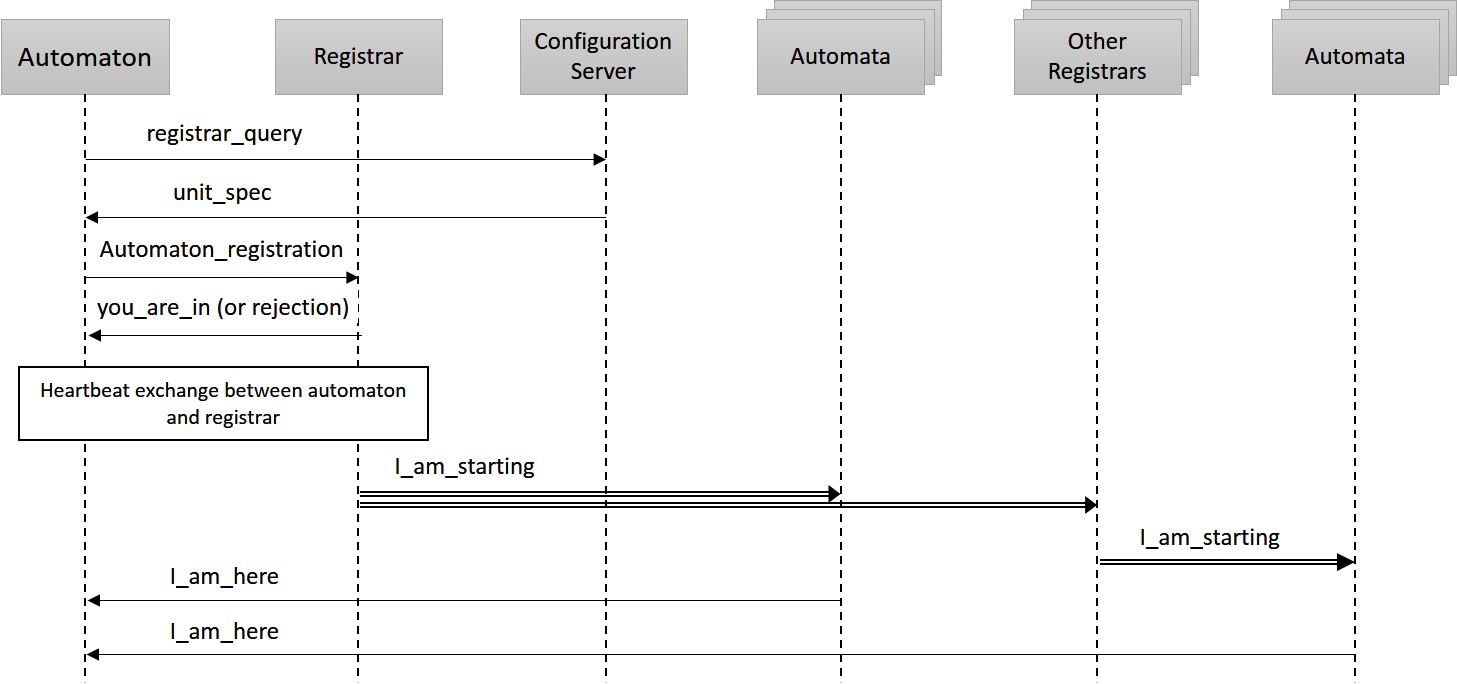} 
  \caption{Automaton registration process}
  \label{fig:automaton-registration}
\end{figure}
As mentioned in Section \ref{sec:distribution}, the distributed nature of medical best practice systems is associated with a high degree of dynamism due to the many uncertainties and condition changes that exist. Given the timely changes in pathophysiological and physical conditions, the automata distribution cutoffs adapt over time, with new automata joining or leaving the system in response to variations in capabilities, patient, or disease models. For example, availability of portable CT scan ambulances \cite{mobileCT} can change the distribution of best practice automata, and introduce new entities with sets of units and automata which overall helps reduce the overall treatment time for stroke patients. Drug complications and side effects or development of side diseases are other examples. For instance, patients treated with antiplatelet agents such as aspirin therapy have a high prevalence of side effects, such as stomach pain, heartburn, or nausea \cite{sideeffect1}. Development of any side effect per se can therefore introduce new disease automata, asking for a dynamic registration approach.

Automata can only operate by identifying themselves to registrars. Automaton registration involves three phases. During the first phase, configuration server is located and identified through repetitive query messages sent to the configuration server at its location address within a specific period of time, which is returned with a ``configuration-server-located'' reply message (location address of configuration is known and pre-defined). Receipt of this message is considered as a successful discovery of configuration server, meaning the location address of configuration server is noted. Otherwise, the procedure is considered not to have succeeded. Availability of a verified configuration server is a necessity otherwise automaton registration would be impossible.

In the next phase, the automaton generates a ``registrar-query'' message with the aim of determining the location address of the automaton's registrar (i.e. registrar of the corresponding unit the automaton belongs to), which is then sent to the configuration server. In case there is no available registrar for the automaton's corresponding unit, a ``registrar-unknown'' message is generated, and is sent back. This failure is updated as a status variable inside the automaton. Otherwise a ``unit-spec'' message is generated and is returned, noting the location address of the corresponding registrar. Once this process is succeeded, the automaton is now allowed to register with its own registrar. A ``automaton-registration'' message is generated and is transmitted to the registrar. The registrar can then reply with two types of messages: a ``you-are-in'' message will be returned if successful, or a ``rejection'' message stating the denial. Figure \ref{fig:automaton-registration} shows the overall process of automaton registration. From this point on, heartbeat messages are being exchanged between the registered unit and the configuration server periodically to monitor unit's availability and to detect registrar's failure.

As for implementation purposes, in case registration was accepted, the registrar generates a ``I-am-starting'' message containing the new automaton's configuration state, which is sent to all other automatons in the unit as well all other registrars to be directed to their respective automatons. Once received, each automaton will reply back with a ``I-am-here'' message containing its unique ID. We consider this approach as the default implementation method. However, other alternatives can be implemented if registrars need to track and keep record of information for their respective automata.

\subsection{Monitoring Failure}

%\begin{figure}[!t]
%  \centering
%  \includegraphics[width=\columnwidth ]{figure/heartbeat.jpg}
%  \caption{Overview of heartbeat message exchange}
%  \label{fig:heartbeat}
%\end{figure}

\begin{figure}[!t]
  \centering
  \includegraphics[width=.9\columnwidth ]{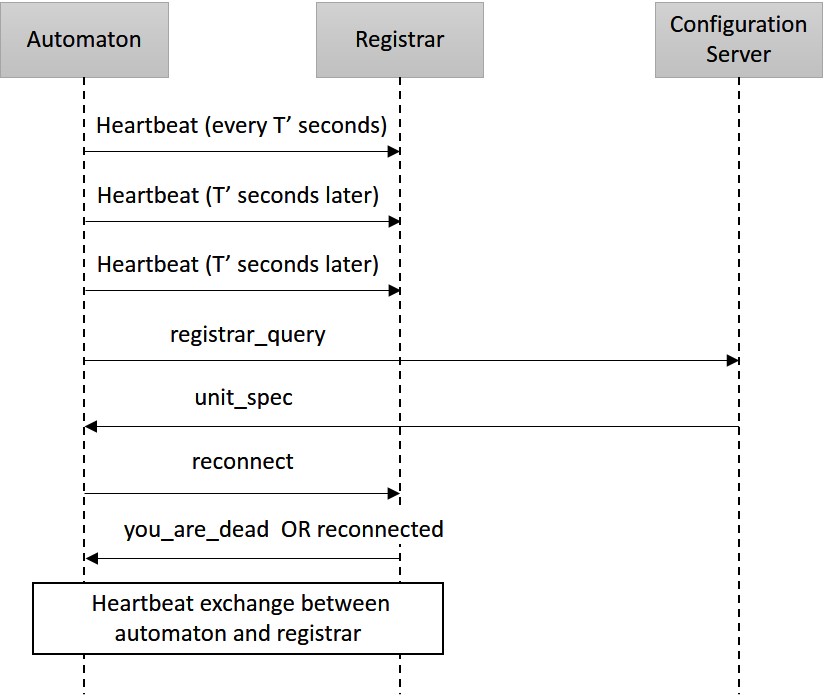}
  \caption{Registrar failure detection process}
  \label{fig:registrar-failure}
\end{figure}
To detect failure and monitor availability of the architectural components, we integrate heartbeat protocol into our architecture. Heartbeat messages are periodically being exchanged between registrars and configuration server, as well as between registrars and their unit's automata, %Figure \ref{fig:heartbeat} provide a visual overview of heartbeat exchanges between an automaton, and the corresponding registrar and configuration server, 
with the time period of heartbeat exchanges between registrar and configuration, and between registrar and automaton set to T and T' seconds, respectively. We use a default heartbeat rate of once every 5 seconds for both T and T', and set N=3 successive missed heartbeats as an indication of termination, as recommended by IEEE Standard 1278 for Distributed Interactive Simulation \cite{ieee1278} and Space Data Systems Standards for messaging services \cite{ccsds1}. %The horizontal dashed lines represent simultaneous messages, and suggest that heartbeat messages are independent events, based on internal conditions or timers. It should also be noted that heartbeat exchanges can either be directly autonomously based on an internal timer, or it can be in response to a required heartbeat receipt.

\subsubsection{Monitoring Health of Automaton}
In order to maintain the availability and also to avoid wasting resources on attempts to send messages to unavailable automata, it is crucial for registrars to keep monitor health, and detect termination of automata they are responsible for. When an automaton terminates, it automatically signals its registrar about its cease of service. However, in case of crashes, or when the automaton is powered off or rebooted, no such signal is being transmitted to the registrar. For this reason, heartbeat messages are periodically being sent from every automaton to its registrar every T' seconds as mentioned earlier.

\begin{figure}[!t]
  \centering
  \includegraphics[width=\columnwidth ]{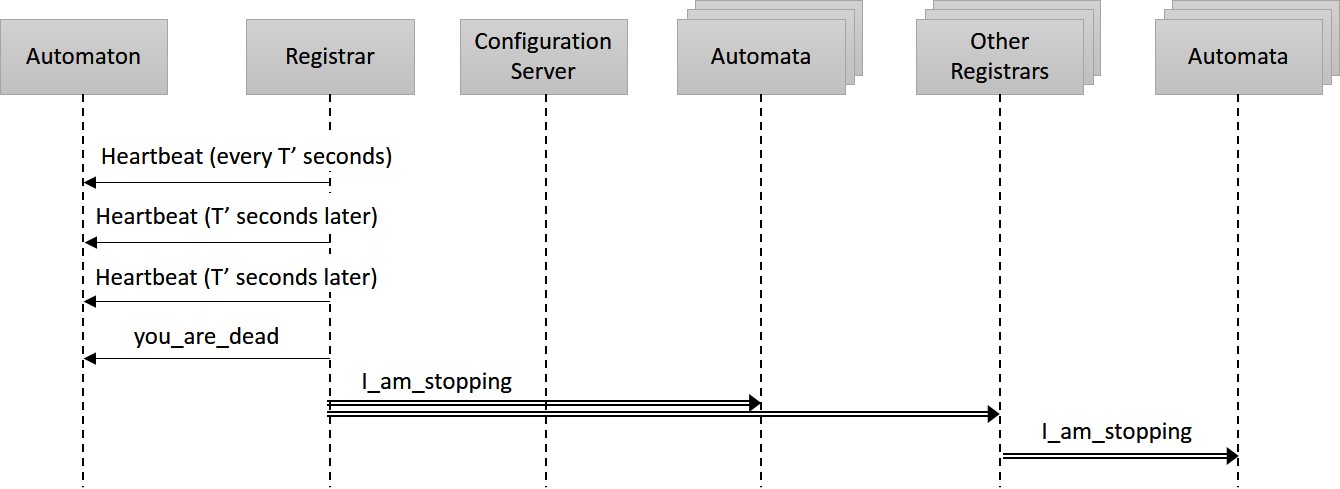} 
  \caption{Automaton failure detection process}
  \label{fig:automata-failure}
\end{figure}

Figure \ref{fig:automata-failure} shows actions taken by the registrar when an automaton failure has been detected. In case of a heartbeat failure, the registrar generates a ``you-are-dead'' message, and sends it back to the automaton, indicating that it is presumed the automaton has failed, is no longer available, or is no longer authenticated. In case the automaton is in fact still running (assume the automaton is hung due to a deadlock such as performing a CPR (Cardiopulmonary Resuscitation) on an unshockable rhythm), the automaton will terminate immediately upon receipt of this message. Following that, a ``I-am-stopping'' message is generated and returned to the registrar, which is then forwarded to all other automatons inside the same unit and other registrars, signaling the termination of the failed automaton.

%NOTE – The value of N4 is defined as twice the value of N3, which is a global implementation parameter within any single entity. If a very large value of N3 is selected, then heartbeat exchange is in effect disabled; that is, the monitoring of module health, registrar health, and configuration server health is optional. AMS entity health monitoring may be unnecessary under certain circumstances, e.g., in tightly controlled spacecraft on-board environments where other mechanisms for the detection of module failure are available. In these environments, the initiation of AMS procedures for handling imputed entity termination—in effect, the simulation of heartbeat failure—is an implementation matter.

\subsubsection{Monitoring Registrar's Health}
In addition to monitoring health of automata by registrars, in a mutual way, every registrars also sends heartbeats to its unit's automatons, so that an automaton can infer its registrar has crashed. When a registrar failure is detected, it is assumed by the automaton that the registrar has been restarted from the time it was failed. In that case, the automaton will query the configuration server to determine the new location address of the restarted registrar and attempts to reconnect. Once reconnected, heartbeat exchanges are resumed. The process is shown in Figure \ref{fig:registrar-failure}.%This presumption is reasonable because the reciprocal heartbeat monitoring relationship between a registrar and its modules is replicated in the relationship between the configuration server and all registrars, but on a slightly shorter cycle. The configuration server interprets N6 successive missing registrar heartbeats as an indication that the registrar has crashed; on detecting such a crash it can automatically take some implementation-specific action to cause the registrar to be restarted, possibly on a different host, so by the time the registrar’s modules detect its demise it should already be running again.
It should be noted that given the automaton heartbeat period, within the first $N\times T'$ seconds after reset, the heartbeat messages from all automatons will be received by the registrar, which therefore helps in the accurate acquisition of unit's configuration.

%NOTE – The worst-case interval between failure and recovery of registration service within a cell is N5 seconds. In the event that N3 is a very large number (disabling heartbeat exchange, as discussed above), implementation-specific mechanisms for simulating the effects of heartbeat failure will reduce this recovery interval to far less than N5 seconds.

%This accurate configuration information must be delivered to new modules at the time they start up (so that they in turn are qualified to orient a newly restarted registrar to the cell’s configuration in the event that the registrar crashes). For this reason, during the first N5 seconds (or other implementation-specific recovery interval, as noted above) after the registrar starts it accepts only MAMS messages from modules that are already registered in the cell (i.e., have been assigned module numbers); if it accepted and processed a registration message from a new module before being certain of the status of all old ones, it would run the risk of delivering incorrect information to the new module.

\subsubsection{Configuration Server Fail-over}
Similar to other components, a configuration server may also presumed to be failed or unavailable, once $N$ successive missed heartbeats are detected by a registrar. In case of this event, the registrar begins cycling through all possible known location addresses for the entity's configuration server, and attempts to re-establish the connection at an alternate location which was thought to be caused by a reboot. During the crash interval, no new automaton can register and get initialized as there is no way of knowing registrar's location due to unavailability of configuration server. The new automatons will also cycle through all known possible location addresses searching for the entity's configuration server to perform initialization procedure. Once the configuration is restarted and was bound to a new location address, all registrars will eventually find it and note themselves to it. Similarly, initialization and registration process of newly joined automata will resume immediately after that.

As mentioned earlier, our architecture supports multi-ranked configuration servers for redundancy purposes, which makes it possible for multiple configuration servers to run concurrently in case one of the configuration servers crashed due to reasons such as a transient network connectivity failure. In that case, every running configuration server periodically sends a ``I-am-running'' message to all lower-ranked configuration servers. Upon receipt of such a message, the respective configuration server stops immediately. That causes all registrars and automata that were communicating with  that configuration server to note its unavailability, which makes them search for the highest-ranked available configuration server. This causes the whole entity get back to service eventually.

\subsection{Open-Loop Safety}
Communication failure in the wireless environment can lead to life-critical safety issues within the message exchange environment. Our message-exchange architecture should guarantee the safety of the execution of distributed best practice automata, to ensure that the automata transit to a safe state even with communication failure or loss of messages. Let's take the stroke automata in Figure \ref{fig:strokeAutomata} as example. Assume a message triggers a state transition event, making the automaton transit to the ``tPA Therapy'' state. Suddenly communication fails, arising the question ``how long to stay in the state and continue tPA therapy?''. Continuing tPA therapy for longer than a specific duration characteristically is hazardous for the patient, therefore considered to be unsafe for the system. Same concept is applied to the "supporting therapy" state using Aspirin as well, which is only allowed for a bounded period of time, given some vital signs changes as per supervising doctor's suggestions. Based on that characteristic, we classify states into the following two classes: 
\begin{itemize}
\item Transient safe state, which allows an automaton to stay safely in the state, but only for a limited duration. That said, if staying on a transient safe state lasts longer than the specified allowed limit, it becomes unsafe, and may lead to hazards. ``tPA Therapy'' state is an example of a transient safe state;
\item An open-loop safe state, which is considered always-safe for the maximum duration of the given medical procedure. Therefore, an open-loop safe state does not involve any hazard while stay lasts more than any time threshold.
\end{itemize}

\begin{table*}[!htb]
\centering
\caption{Message header fields (64-bit total)}
\label{table:header}
\begin{tabular}{|l|l|l|}
\hline
\multicolumn{1}{|c|}{{\bf Field}} & \multicolumn{1}{|c|}{{\bf Length (bits)}} & \multicolumn{1}{|c|}{{\bf Description}}                                                                          \\ \hline
Message type                      & 6                                        & UID representing type of message               \\ 
\hline
Priority                          & 3                                        & 0-7: A value representing the urgency of message  \\ 
\hline
Checksum flag                     & 1                                        & Value is set to 1 if the data is followed by a 32-bit checksum. %Set to 0 if no checksum exists 
\\ 
\hline
Open-loop safe state              & 8                                        & UID representing the next safe state in case of communication failure\\
\hline
Source entity number         	  & 5                                        & UID representing source entity                                                                               \\ 
\hline
Source unit number                & 5                                        & UID representing source unit                                                                                    \\ \hline
Source automaton number           & 5                                        & UID representing source automaton                                                                        	    \\ \hline
Destination entity number   	  & 5                                        & UID representing destination entity                                                                          \\ 
\hline
Destination unit number           & 5                                        & UID representing destination unit                                                                               \\ \hline
Destination automaton number      & 5                                        & UID representing destination automaton                                                                           \\ \hline
Application-specific data length  & 16                                       & Length of application-specific data (limited to 65,000 bits)           							   \\ 
\hline
\end{tabular}
\end{table*}

To maintain reliability and safety, our designed communication protocol must ensure open-loop safety to guarantee that the system transits from a transient safe state to a predefined open-loop safe state in case a communication failure occurs. Therefore, we embed open-loop safety as a safety parameter into our protocol, so that a message triggering a state transition forces the automaton not to make a state transition unless an open-loop state is determined, and queued as an emergency option in case communication fails. Given the stroke example, possible transient safe states such as ``tPA Therapy'' are transited to an \textit{implicit} ``general assessment'' state as an open-loop safe state, so to ensure the safety requirements of any automaton.

\subsection{Remote Message Exchange}
\label{RME}
Automata in different entities exchange messages through Remote Message Exchange (RME) gateways. The RME gateways have access to other entities' RME gateways through establishment of a persistent connection using their network interfaces, that in overall form a tree of mutually aware interconnected entities for the distributed best practice systems, enabling a message to get forwarded to any desired automaton placed at any distributed location. Upon receipt of a message, the RME gateway forwards the message to any number of target automata. Using a publish-subscribe messaging model can therefore help as the copies can be published to only the subscribed automata. The protocol is efficient in the sense that only a single copy of messages is ever being sent over the link, no matter how many automata intend to receive copies of the messages. The use of RME gateways cannot insulate the effect of latency variation on message propagation. This fact is assumed to be be handled by the underlying communication layers, which is beyond design of our higher-layer message-oriented architecture.

\subsection{Communication Protocol Header}
\label{header}

Our pathophysiological communication protocol consists of a header in fixed format as a 64-bit prefix to a packet, which is followed by zero or more octets of data, plus an optional 32-bit checksum. Table \ref{table:header} shows an overview of our message header fields. The first header field in a message is the 6-bit message type, which represents the type of the message that is being exchanged, either application-specific data messages (e.g. a neurological symptom, disease state, patient state confirmation, time log, etc.), or configuration messages (e.g. heartbeat, query, acknowledgement messages, etc.). The priority field is a 3-bit field indicating a value 0 to 7, which represents the urgency of the message, with higher values representing higher priority of messages. Consideration of the priority field is inspired by the fact that the urgency of messages are in fact state-dependent, which requires situation awareness~\cite{situationAwareness}. Given the case of our stroke scenario for example, at the time of assessing a suspected stroke to detect whether the type of stroke is ischemic or hemorrhage, transmission of lab results such as TEG values for blood coagulation level has a slightly higher priority than general vital signs such as heartbeat or blood pressure, while both have higher priorities than video data for remote screening of patient. A checksum may optionally be provided as the last 32 bits of any messages, using the ISO/IEC 3309 -compliant 32 bit CRC algorithm \cite{ISO3309}. This algorithm is also compliant with the frame checking sequence as defined in section 4.2.5.3 of the ISO/IEC 13239 specification \cite{ISO13239} and section 8.1.1.6.2 of ITU-T recommendation V.42 \cite{ITU}. The presence of a checksum is indicated by a set value of ``1'' in the checksum flag field, while otherwise is set to ``0''. To ensure open-loop safety, our protocol header reserves a field for open-loop safe state, which represents the UID of a safe state that must be perpetuated as a permanent safe state where the automaton must transit to in case of undesirable unsafe situations or communication failure. Next fields represent source and destination entity number, unit number, and automaton number storing the hierarchical source and destination address, respectively, used for end to end forwarding of messages. As the length of the pathophysiological data in a message varies, the data length field of the header indicates the length of data which is followed, limited to 65,000 bits.

\section{Proof-of-Concept Simulation}
\label{sec:implementation}

%\begin{figure}[!t]
%\centering
%\includegraphics[width=1\columnwidth]{figure/sepsis.png}
%\caption{Simplified statechart model for Sepsis.}
%\vspace{-.4cm}
%\label{Fig:sepsis}
%\end{figure}
\begin{figure*}[!t]
\centering
\includegraphics[width=1\textwidth]{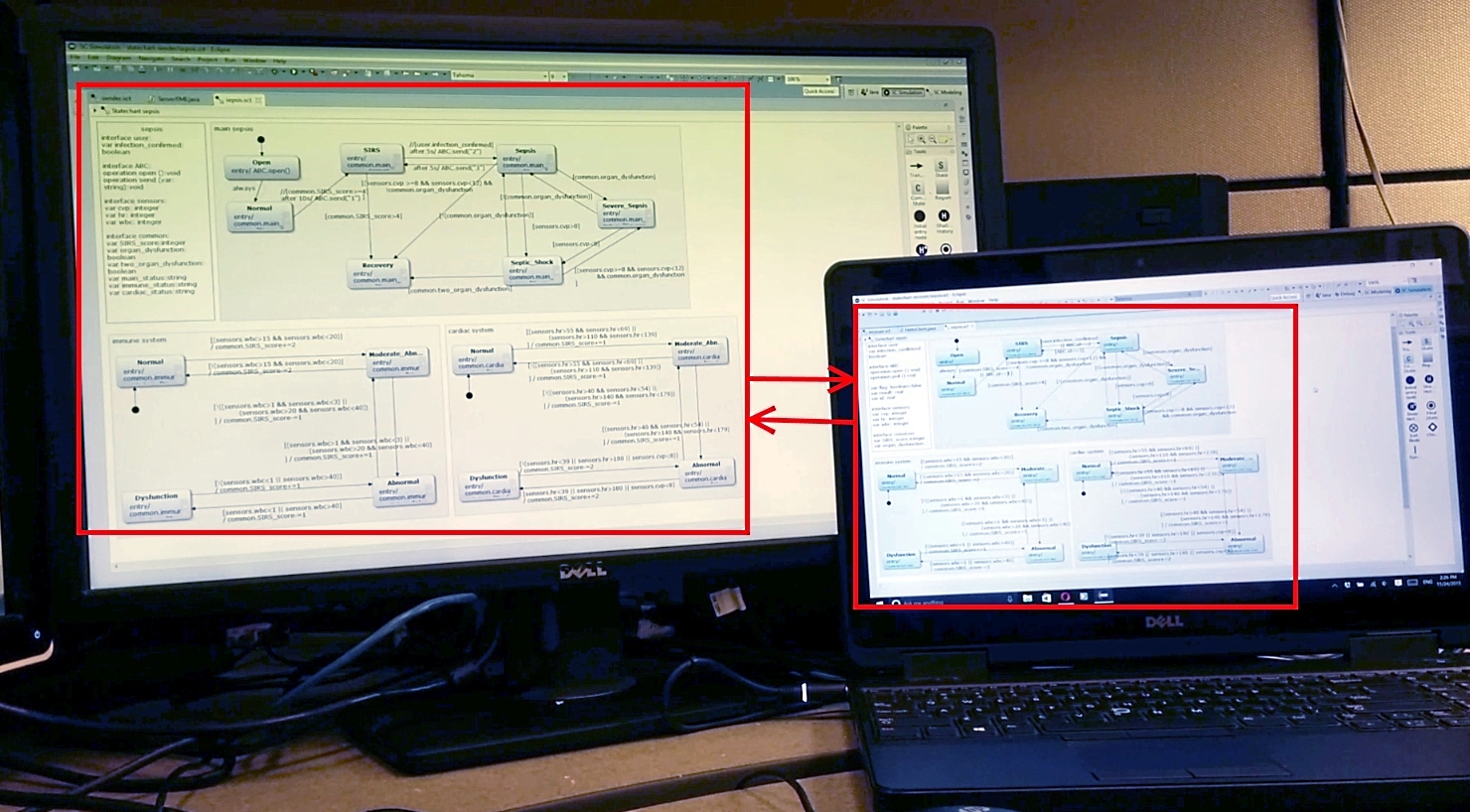}
\caption{A sample of real platform experiments.}
%\vspace{-.4cm}
\label{Fig:experiments}
\end{figure*}

\begin{figure*}[htbp]
    \centering
        \includegraphics[trim=.75in 4.3in .75in 4.3in, width=1\textwidth]{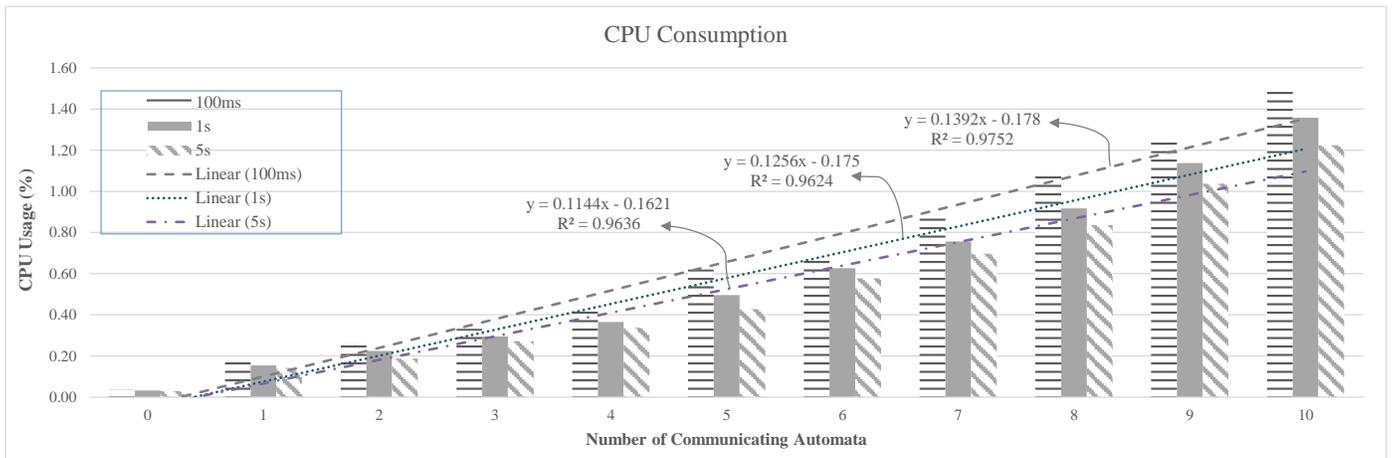}
    \caption{The performance overhead of our message-exchange system in terms of CPU usage.}
    \label{Fig:cpu}
\end{figure*}

We have developed best practice medical executable automata as proof-of-concept case studies, and implemented and tested our message-exchange communication architecture rigorously over these case studies conducted in collaboration with Carle Foundation Hospital~\cite{carle}. We performed our experimentations on a real platform where 230 synchronization requirements were specified to synchronize two sets of distributed medical automata. To develop our best practice medical automata, we used Yakindu statechart 2.4 open-source tool on top of Eclipse Luna 4.4.0 IDE to model the automata as executable statechart models which can further enable rapid prototyping and validation with domain experts \cite{yakindu}. Our developed best-practice statechart models include the simplified \textit{stroke} as well as simplified \textit{sepsis} medical best-practice automata consisting of both disease and underlying organ models, all represented as executable statecharts. The models focus on adherence to best-practice medical guidelines, which are codified from medical knowledge, simplified, and then validated with physicians for correctness. A pair of each set of statechart models are distributed on two different machines- one set representing rural hospital or ambulance, and the other representing center hospital. We implemented the messaging communication system in Java, so that it can be deployed on any platform running Java Virtual Machine (JVM), including Linux and Windows. We have designed a list of APIs for the users, including establishment of a connection, composing messages, and message pushing and polling operations. Figure \ref{Fig:experiments} illustrates a sample of our experiments \footnote{A demo illustrating a part of our simulation is available at:\\ http://publish.illinois.edu/mdpnp-architecture/672-2}.

As our communication architecture supports different message-oriented patterns and any combinations of these patterns to meet application-specific requirements, we employ a push-poll pattern to implement our model-driven communication system, consisted of a push client and a polling client to form a registrar, as well as a synchronized FIFO queuing module, all residing on each statechart machine. The messages are captured by the push client and are encoded to a specific message format as defined per our communication protocol described in Section \ref{header}. The messages are then encrypted with the AES 128-bit symmetric cipher in electronic codebook (ECB) mode, serialized, buffered, and are eventually pushed into the queuing module via a persistent socket connection to be polled by corresponding destination registrar.

While sending is good for individual distributed statecharts to push messages to the synchronized queuing module, this is not yet a good approach for distributing the messages among distributed models as it imposes a significant overhead on the registrars. The registrars are forced to inefficiently keep a long-lived and mostly unused network socket connection open with eachother. This leads us to use client-side polling for registrars, for two main reasons: First, client-side polling is architecturally simpler. Using this approach, the registrars doesn't have to track which registrars called and which registrars are waiting for replies. This leads to simpler implementations while also making it easier for supporting various types of registrars. Nevertheless, the most efficient option here is to poll values in a guard expression. We set the polling frequencies to 200ms as a result of trade-offs between callback frequencies and processing overhead, which meets the real-time synchronization requirements of our hospital setting. 

To evaluate the performance of our message-exchange architecture, we profiled the CPU usage of our system, and instrumented the CPU consumption of all the underlying threads. This is useful especially for identifying components that have high CPU consumption which also can be clues of deadlocks. During the profiling sessions, no abnormal CPU usage was detected by any specific thread. Figure \ref{Fig:cpu} shows average CPU usage for various number of communicating automata (baseline case with no communicating automaton, up to 10 concurrent communicating automata), for three different polling rates (100ms, 1s, and 5s). As can be seen, overall, the overhead of our message-exchange system is negligible, and no sudden spike can be noticed in the load. The average CPU consumption for 10 communicating automata with polling rates of 1s are less than 5\%. The low CPU utilization of our message-exchange system also signifies that no source code problems such as infinite loops or excessive backend calls, and no excessive garbage collection cycles take place inside the runtime execution of the system. The limited number of active threads in our message-exchange system helps with lowering the CPU consumption and the overall performance as the number of context switches are also limited. Figure \ref{Fig:cpu} also illustrates that the CPU consumption almost follows a linear-like trendline with high confidence (\texttt{R-Squared} value of more than 95\% fitting the linear regression lines), therefore making our message-exchange system scalable in terms of number of communicating automata. Interestingly, the small difference in the slope of linear regression trendlines indicates that the performance overhead of our message-exchange system is not significantly influenced by the polling rates of registrars, therefore making our message-exchange architecture more robust in higher polling rates.

The communication and synchronization requirements were inspected multiple times with multi-disciplinary domain experts (10 developers, 12 researchers, and 4 physicians) to ensure that specific functional and medical requirements were satisfied and accomplished correctly. Apart from the real experiments and the important benefits resulting from using our message-exchange architecture, we have received positive feedback from the experts witnessing our message-exchange architecture. The qualitative feedback we received is promising and suggests that the middleware can in fact be applicable to large sets of requirements and that it can be extended to domains that than medical services. Such domains include large-scale co-simulation of heterogeneous production and ERP software models especially in the automotive industry~\cite{automotive}.

\section{Conclusion and Future Work}
\label{sec:conclusion}
In this paper, we describe a dynamic distributed executable medical best practice guidance system. The design provides a platform to assist adherence to medical best practices in locations throughout a distributed healthcare provider network: from rural hospital, through ambulance transfer, to regional tertiary hospital center. We codified complex medical knowledge into simplified executable automata, and targeted stroke as the case study demonstration to illustrate and motivate the  synchronization of distributed medical best practice models. The design is founded on a dynamic pathophysiological model-driven message-exchange architecture; our proposed message-exchange architecture meets the dynamism, safety, and reliability requirements for communication and synchronization across distributed emergency medical best practice systems. We implemented the communication system and applied it using proof-of-concept medical best practice automata. Stroke model medical best practice simulations were conducted.

In the future, we plan to clinically validate the communication system in collaboration with Carle Foundation Hospital \cite{carle}, run extensive performance assessment, and implement it on a real clinical testbed that we have built using SimMan medical patient simulator \cite{simman}. We also intend to systematically evaluate our communication system using quantitative metrics, and formally verify the protocol to make sure it is always safe for all random combination of inputs.
%\newpage
\bibliographystyle{IEEEtran}
\bibliography{ref}

\end{document}